\documentclass{aa}  
\usepackage{graphicx}
\usepackage{txfonts}
\begin{document}

\title{Disentangling the morpho-kinematic properties of  a \\ face-on meger at z$\sim$0.7}
\author{I. Fuentes-Carrera\inst{1,2}
          \and 
           H. Flores\inst{1}
          \and
          Y. Yang\inst{1,3} 
	  \and
          S. Peirani\inst{4}  
          \and
           F. Hammer\inst{1}
          \and
           M. Rodrigues\inst{1,5}
          \and
           C. Balkowski\inst{1}   
          }

\offprints{I.Fuentes-Carrera}

\institute{GEPI, Observatoire de Paris, UMR 8111 CNRS, Universit\'e Paris Diderot, 
5 Place Jules Janssen, 92190 Meudon, France 
\and
{Escuela Superior de F\'\i sica y Matem\'aticas, Instituto Polit\'ecnico Nacional (ESFM-IPN), U.P. Adolfo L\'opez Mateos, edifico 9, Zacatenco, 07730, Mexico City, Mexico}  \\
\email{isaura.fuentes@obspm.fr,isaura@esfm.ipn.mx} 
\and
{National Astronomical Observatories, Chinese Academy of Sciences, 20A Datun Road, Chaoyang District, Beijing 100012, China}   
\and
Institut d'Astrophysique de Paris, UMR 7095 CNRS, Universit\'e Pierre et Marie Curie, 98 bis Bd Arago, 75014 Paris, France 
\and
CENTRA, Instituto Superior T\'ecnico, Av. Rovisco Pais 1049-001 Lisboa, Portugal \\
  }
   \date{ }

 
  \abstract
   {Galaxies with masses between 10$^{10}$ and 10$^{11}$ solar masses have suffered significant evolution since  z$\sim$1. 
At intermediate redshifts, many galaxies seem to be perturbed or suffering from an interaction.
It is not always straight-forward to determine what type of encounter or perturbation is observed, nor the outcome of such an event.
Considering that disk galaxies may have formed and evolved through minor mergers or through major mergers, it is important to understand the mechanisms at play during each type of merger in order to be able to establish the outcome of such an event. 
In some cases, only the use of both morphological and kinematical information can disentangle the actual configuration of an encounter at intermediate redshift.
}
   {In this work, we present the morphological and kinematical analysis of a system at z=0.74 in order to understand its configuration, interacting stage and evolution.}
   {Using the integral field spectrograph GIRAFFE, long-slit spectroscopy by FORS2 and direct optical images from the HST Advanced Camera for Surveys and ISAAC near-infrared images, we disentangle the morphology  of this system, its star-formation history and its extended kinematics in order to propose a possible configuration for the system. Numerical simulations are used to test different interacting scenarii. }
   {We identify this system as being a face-on disk galaxy with a very bright bar  in 
interaction with a  smaller companion with a mass ratio of 3:1. The relevance of kinematical information and the constraints it imposes on the interpretation of the observations distant galaxies is particularly strengthened in this case. 
}
   {This object is amongst the best example on how one may misinterpret morphology in the absence of kinematical information.}


   \maketitle
%

\section{Introduction}

Understanding how and when galaxies formed is one fundamental open question in astrophysics.
To address the problem of galaxy evolution, one needs to understand how high-redshift galaxies evolve into the galaxies seen at low redshift. 
Galaxy formation and evolution in a cold dark matter (CDM) dominated universe can be described as follows: in the early universe, gas collapsed in dark matter halos, which later cooled to form stars, creating the first galaxies. Later these dark halos merged to form larger dark halos and thus more massive galaxies (e.g., Cole et al. 2000; Somerville, Primack, \& Faber 2001). In this hierarchical picture, galaxy interactions and mergers are an essential ingredient. 
Earlier numerical simulations have shown that disks formed around previously existing spheroids through the smooth accretion of  gas from the intergalactic medium (e.g., Steinmetz \& Navarro 2002), whereas the spheroids are the remnants of major merger events where disks are thrown together and mixed violently on a short timescale (e.g., Toomre 1977, Barnes \& Hernquist 1992, Mihos \& Hernquist 1994). According to this scenario, disks are fragile: they are destroyed in major mergers, and they are heated in minor mergers. 
However, more recent simulations (Barnes 2002, Springel \& Hernquist 2005) have shown that gas-rich mergers can make new disks. These results support the spiral rebuilding scenario proposed by  Hammer et al. (2005) in which  mergers can also be responsible for the formation of disks. 
This scenario renamed ``disk survival'' is also proposed by Hopkins et al. (2009).

We would like then to identify the types of merging episodes that end up as a disk galaxy.
In order to do so, a joint analysis of morphological information with kinematical studies and spectral information is needed. The complementarity of these different types of analysis is essential to properly characterize distant galaxies  in order to avoid any misinterpretations due to the lack of resolution at larger redshifts. 
The IMAGES (Intermediate MAss Galaxy Evolution Sequence -Yang et al. 2008) program is an ESO Large Program exploiting data from large ground telescopes and from space for such a purpose. IMAGES makes use of 1D and 2D spectroscopical data and direct imaging in the optical and the near-infrared (NIR) to shed more light on the following questions:
How did galaxies form and assemble their stellar mass? When was the morphological differentiation of galaxies established? How did the Hubble Sequence form?

In this work we present the morphological and kinematical analysis of  J033227.07-274404.7, a system at z$\sim$ 0.74 displaying an unusually elongated, arc-like morphology with a velocity field that does not seem to match the observed morphology .
Section  \ref{data} presents the overall properties of the object and the available data. In Section \ref{analysis} the analysis of the data is presented, while in Section \ref{conf}, we present a the possible configuration of the system, as well as a numerical model of a probable encounter. Finally, in Section \ref{disc} we discuss the results and present the conclusions of this work.

In this paper we adopt the following cosmological parameters: $H_0$ = 70 km/s/Mpc,  $\Omega_M = 0.3$ and $\Omega_\Lambda= 0.7$. Magnitudes are in the AB-system.

\section{Data}
\label{data}

J033227.07-274404.7 is an object in the Chandra Deep Field South (CDFS).
High spatial resolution images of the system were taken from the HST archive. The images were obtained with the Advanced Camera for Surveys (ACS) in its  Wide Field Channel (WFC)  mode on the HST as part of the GOODS projects (Giavalisco et al. 2004). Images were taken in the following filters: F435W, F606W, F775W and F850LP (Beckwith et al. 2006).
These filters are close to the B, V, i and z passbands, respectively. The publicly available version v1.0 of the reduced, calibrated, stacked, mosaicked and drizzled images were used (drizzled pixel scale = 0.03 arcsec). 
Ground-based images in the U, B, V, R and I bands were taken from the ESO Imaging Survey (EIS -Arnouts et al. 2001). These images were taken with the WFI camera on the MPG/ESO 2.2m telescope. Each image has a pixel size of 0.238 arcsec with an average seeing of 0.86 arcsec.  
Images in the J, H and Ks bands were taken from EIS deep survey (Vandame et al. 2001). These images were taken with ISAAC with a plate scale of 0.15 arcsec/pix with a reported seeing of 0.6 arcsec.
The object is detected by Spitzer/IRAC in the mid-infrared (Dickinson et al.-in preparation) and with Spitzer/MIPS in the far-infrared (Chary et al.-in preparation). However the object shows almost no emission in these wavelengths and is greatly contaminated by a southern galaxy at z$>$1. The limiting magnitudes for these observations were 21.3 mag at 24$\mu$m for MIPS, and 26.15 mag, 25.66 mag, 23.79 mag and 23.70 mag for IRAC's channels one to four, respectively.
The object was not detected by GALEX in the FUV (most likely due to the large redshift of the object). In the NUV, two sources are detected at less than 2 arcsec of the object. However, considering the pixel size of GALEX (1.5 arcsec), these detections might be contaminated by the presence of the neighboring galaxies -especially the galaxy at z$\sim$0.63. 
The object was not detected in CHANDRA observations down to 5.5 x 10$^{17}$ ergs cm$^{-2}$ s$^{-1}$ in the 0.5-2 keV band and 4.5 x 10$^{16}$ ergs cm$^{-2}$ s$^{-1}$ in the 2-10 keV band (Giacconi et al. 2002; Rosati et al. 2002).

\begin{figure*}
\centering     
\includegraphics[scale=0.33, angle=0]{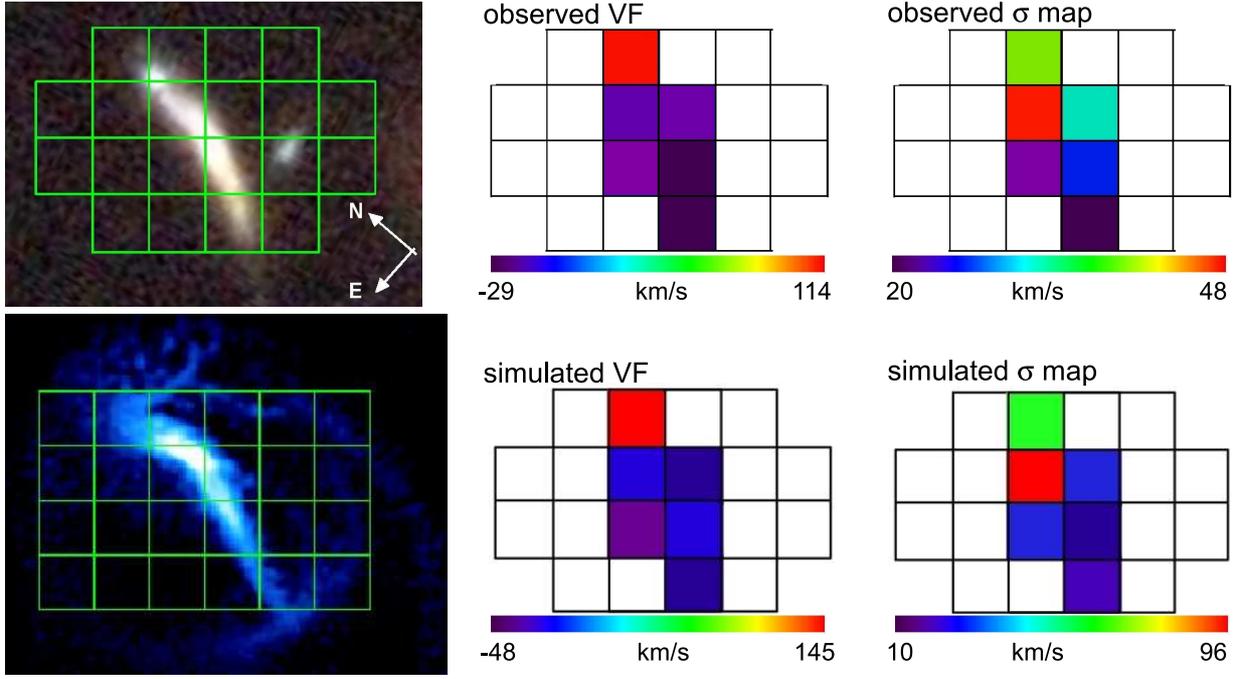}
\caption{{\it Top row, left panel}:  Composite image (B+V, i and z bands) of J033227.07-274404.7 taken with the Advanced Camera for Surveys (ACS) of the HST. 
Size of FoV is 2.0 arcsec $\times$ 2.6 arcsec (14.4 kpc $\times$ 18.8 kpc). Superposed grid shows the position of the GIRAFFE IFU.
{\it Top row, middle panel:}  Line-of-sight velocity map of the object 
derived from GIRAFFE observations of the [OII]3726,3729 \AA \ doublet.
{\it Top row, right panel:} velocity dispersion, $\sigma$, map of the object 
derived also from GIRAFFE observations.
{\it Bottom row, left panel}:  Projected distribution of gas of the simulated encounter using the ZENO code (Barnes 2002). Snapshot shows the encounter just after the second peripassage (12 Myr after)  and 1.93 Gyr after the first peripassage. The FLAMES IFU grid is superposed to simulate the GIRAFFE observations. 
{\it Bottom row, middle panel:}   Line-of-sight velocity field derived from the ZENO simulation considering an IFU similar to that of GIRAFFE.
{\it Bottom row, right panel:} $\sigma$ map of the same simulation 
}
\label{objeto_sims} 
\end{figure*}

For each image, photometry of the system was extracted within a constant aperture of 3 arcsec in diameter. 
Because of the large uncertainty associated with the zero point calibration of the H-band, we used the revised calibration proposed by Wuyts et al. (2008).
Table \ref{phot} shows the photometry values and associated uncertainties for J033227.07-274404.7 
Uncertainties in the extracted magnitudes correspond to the magnitude error as given by Sextractor (Bertin \& Arnouts, 1996), and the systematic error associated with the uncertainty on the zero-point (ZP) calibration. 
Due to the weakness of the emission for $\lambda > 35.0 \mu m $ and the contamination from the galaxy at larger redshift, no photometry was extracted at these wavelengths.

Two-dimensional spectroscopic observations of J033227.07-274404.7 were taken with the integral field spectrograph FLAMES (Fibre Large Array Multi Element Spectrograph) with the medium-high (R=5600-46000) resolution spectrograph GIRAFFE (multi-IFU mode) as part of the IMAGES program (Yang et al. 2008).
The object was observed with one IFU consisting of a rectangular array of 20 microlenses of 0.52 arcsec each, giving an aperture of 2 arcsec $\times$ 3 arcsec. The LR05 setup was used to observe the redshifted [OII]3726,3729 \AA \  doublet. The line-of-sight velocity and velocity dispersion at each lenslet were derived following the procedure described in Flores et al. (2006).

A low-resolution (R=860), large wavelength-range long-slit spectrum (6000 \AA \ to 10800 \AA) 
of the object was taken with the visual and near UV Focal Reducer and low dispersion Spectrograph, FORS2 as part of the ESO-GOODS program of spectroscopy of faint galaxies in the CDFS (Vanzella et al. 2005). 
FORS2 was used in its multi-object spectroscopy with exchangable masks (MXU) mode. The 300I grism was used to obtain a spectral resolution of 3.2 \AA \ per pixel  and a spatial scale plate of 0.126 arcsec/pix.

\section{Analysis}
\label{analysis}

\subsection{General properties}
\label{props}

Top left panel of Figure \ref{objeto_sims} shows the HST-ACS composite image of the object (F435W+F606W, F775W and F850LP filters). The object displays an 
arc-like shape 2.04  arcsec   long (14.3 kpc at a the assumed redshift).
The object seems to be composed of an elongated structure of about 1.52 arcsec (11.0 kpc) long and a smaller structure, 0.34  arcsec (2.45 kpc) long,  located to the north-northwest of the main structure. The inner parts of the bigger structure are brighter than the rest of the structure; fainter, distorted emission is seen at each end of this elongated structure. Small discontinuities in the brightness of this structure are also seen. 
The smaller, northern, structure seems almost circular. 
 A small faint structure is also seen to the south-west of the elongated structure. 

 Spectroscopic observations indicate that the object is at a redshift of 0.73814   with an uncertainty of 5 km/s (Puech et al. 2008). It is surrounded by various objects. A disk-like galaxy to the south-west with a spectroscopic redshift of $1.128$ and another galaxy-like object to the south-east with a photometric redshift of $0.63$.

\begin{figure}
\centering
\includegraphics[scale=0.25, angle=0]{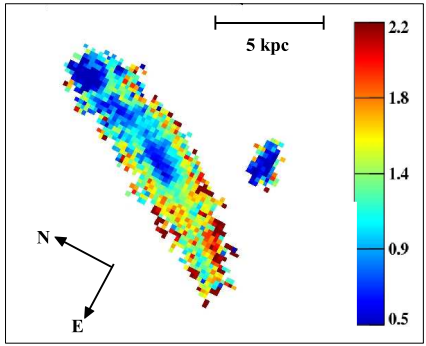} 
\caption{(B-z) color image of J033227.07-274404.7 derived from the HST-ACS images.}
\label{color_objeto} 
\end{figure}

Figure \ref{color_objeto} shows the (B-z) color image of the system. 
This bluest-minus-reddest band image shows that the northern part of the system is much bluer than the southern parts. The region corresponding to the northern smaller structure is the bluest one. 
Bluer colors are seen all the way from the center of the main galaxy to the companion. The smaller region to the south-west also displays very blue colors.
It is important to notice however that in this image even the reddest colors correspond to a (B-z) value lower than 2.7 mag. Following the analysis of Neichel et al. (2008), for a pure starburst region, the observed (B-z) of a galaxy at this redshift is lower than 1.0 mag. A (B-z) value of 2.8 corresponds to a Sbc galaxy at the redshift of the object. So in general, the whole galaxy is forming stars, more importantly in the bluer regions.
Dust has little influence on the color map since, as it will be shown later (Section \ref{SED}), attenuation is low.

\subsection{Kinematics}
\label{kins}

Top middle panel of Figure \ref{objeto_sims} shows the line-of-sight velocity field (VF) of the object. 
Only pixels with a signal-to-noise (S/N) value over 3 were taken into account to derive the VF. The VF encompasses 2.33 arcsec ($\sim$ 16.7 kpc).
Top left panel of Figure \ref{objeto_sims} shows the superposition of the IFU on the composite image of the system.
The pixels corresponding to the elongated structure show a very uniform VF. For these pixels, the global velocity gradient is less than 14 km/s.  On the other hand, the upper northern-most pixel has a velocity difference of more than 130 km/s with respect to the other pixels. 
The position of this pixel matches the location of the smaller, rounder structure to the north-northwest of the main structure.
Top right panel of Figure \ref{objeto_sims} shows the velocity dispersion ($\sigma$) map of the object. 
Values for $\sigma$ go from 20 km/s to almost 50 km/s.
The pixel with the maximum value for the dispersion, $\sigma$=48 km/s, corresponds to the region between the elongated structure and the smaller structure to the north-northwest.
Median seeing during the observation of this object was 0.805 arcsec. 
Considering the large size of each GIRAFFE spaxel (0.52 arcsec), the resulting velocity and dispersion velocity maps are not greatly affected by this value.

\begin{table*}
\caption[]{Fluxes for  J033227.07-274404.7 within 3.0 arcsec. Last column shows the extracted fluxes for the companion and for the central parts of the main galaxy using POLYPHOT.}
\centering
\begin{tabular}{l c c c c c c}
\hline
Filter       & & Magnitude $\pm$ error & & Magnitude $\pm$ error & & Magnitude $\pm$ error \\
             & & within 3 arcsec       & &  main galaxy          & & companion       \\
\hline
U'$_{EIS}$    & & 22.7826 $\pm$ 0.0395  & & --- & & --- \\
U$_{EIS}$    & & 22.8260 $\pm$ 0.0395  & & --- & & --- \\
B$_{ACS}$    & & 23.3267 $\pm$ 0.0194  & & 25.280 $\pm$ 1.180    & & 23.830 $\pm$ 0.605  \\
B$_{EIS}$    & & 23.5340 $\pm$ 0.0165  & & --- & & --- \\
V$_{EIS}$    & & 23.2758 $\pm$ 0.0419  & & --- & & --- \\
V$_{ACS}$    & & 22.9400 $\pm$ 0.0118  & & 24.396 $\pm$ 0.732    & & 22.706 $\pm$ 0.336 \\
R$_{EIS}$    & & 22.5642 $\pm$ 0.0248  & & --- & & --- \\
i$_{ACS}$    & & 22.3859 $\pm$ 0.0137  & & 24.689 $\pm$ 0.965    & & 22.744 $\pm$ 0.394 \\
I$_{EIS}$    & & 21.9494 $\pm$ 0.0338  & & --- & & --- \\
z$_{ACS}$    & & 22.1979 $\pm$ 0.0141  & & 24.769 $\pm$ 1.295    & & 22.898 $\pm$ 0.597 \\
J$_{ISAAC}$  & & 22.09   $\pm$ 0.02    & & --- & & --- \\ 
H$_{ISAAC}$  & & 22.13   $\pm$ 0.03    & & --- & & --- \\ 
Ks$_{ISAAC}$ & & 21.61   $\pm$ 0.02    & & --- & & --- \\  
\hline
\end{tabular}
\label{phot}
\end{table*}

\subsection{Spectral energy distribution} 
\label{SED}

\begin{figure}
\centering      
\includegraphics[scale=0.3,angle=0]{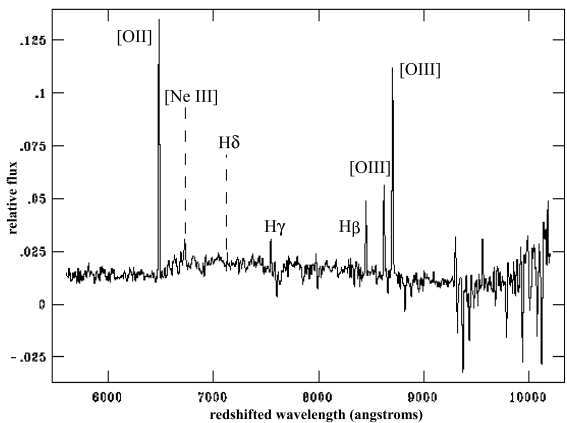}
\caption{FORS2 spectrum of J033227.07-274404.7 Redshifted wavelengths are shown. Relative flux is in energy per area unit per time unit per wavelength. 
}
\label{fors2_spec} 
\end{figure}

The FORS2 spectrum of the system is shown in Figure \ref{fors2_spec}. The [OII]3726,3729 \AA \ doublet is easily identified.  H$\beta$, H$\gamma$ and H$\delta$ are seen in emission. The [OIII]4959,5007 \AA \ lines are also visible. 
We have extracted some properties of the gas phase following the methodology discribed in Rodrigues et al. (2008). Since H$\gamma$/H$\beta$=0.4, there  is apparently no extinction. No evidence for AGN contamination has been found. Indeed, the log[OII]/$\beta$=0.55 and log[OIII]/$\beta$=0.6 ratios are concordant with the expected values of a HII region according to Mc Call et al. (1985). However, the  [NeIII]3869 Å  line is very strong and can be a signature of shocks. The metallicity of the gas is subsolar, with 12 + log O/H= 8.48 dex. This metallicity is tipical of galaxies of mass logM*=10.00 dex at z$\sim$0.7 according to the stellar mass-metallicity relation (Rodrigues et al. 2008).
In order to have a first constraint of the stellar population content of the galaxy, we have performed a fit of the  observed spectrum with a linear combination of stellar libraries.  We used a base of 39 templates single stellar population (SSP) from Bruzual \& Charlot (2007), with 13 ages (from 10 Myr to 5 Gyr) and 3 metallicities (subsolar, solar and supersolar) using STARLIGHT sotfware (Cid-Fernandes et al. 2005). A Cardelli et al. (1989) extinction law was assumed. The best fit shows that a high fraction of the galaxy spectra is domitated by young star features typical of 13 Myr, 100 Myr, 200 Myr and intermediate age stellar population of around 1 Gyr. However the quality of the spectra, low S/N and limited wavelenght range, does not allow us to provide strong constraints on the fractional contribution of light from stellar populations of different ages.
In order to have better constrain in the stellar population, we have investigated the spectral energy distribution (SED) based on photometric measurements.

\begin{figure}
\centering
\includegraphics[scale=0.5, angle=0]{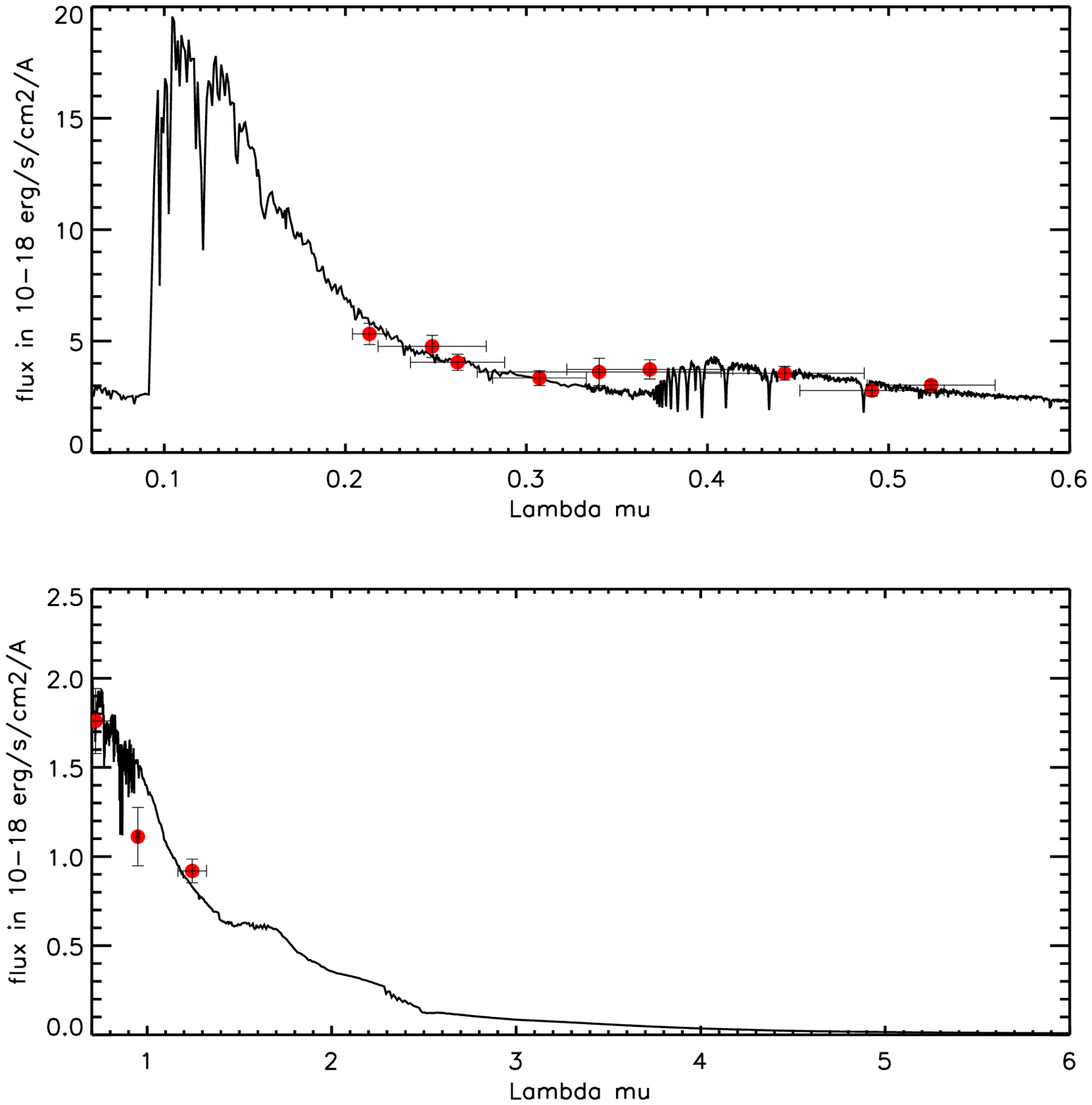} 
\caption{{\it Top panel}:  Spectral energy distribution (SED) of J033227.07-274404.7 in the optical within a 3.0 arcsec aperture.  {\it Bottom panel:}  SED of J033227.07-274404.7 in the NIR. For both SEDs, the wavelengths shown are the rest-frame wavelengths.
 A SED built from stellar synthesis models by Bruzual \& Charlot (2007) have been fitted to the object.}
\label{sed_objeto} 
\end{figure}

Figure \ref{sed_objeto} shows the SED of the whole system using the photometry values presented in Section \ref{data}. Wavelengths shown are rest-frame wavelengths.  Flux is given in ergs/s/arcsec$^2$/angstrom. For the optical part of the distribution, both the ACS-HST and EIS images were used. Full horizontal error-bars represent the FWHM of the photometric filters used in the corresponding surveys, while vertical error-bars represent the convolution of two terms: the magnitude error as given by Sextractor (Bertin \& Arnouts 1996), and the systematic error associated with the uncertainty on the zero point calibration. The resulting SED shows a global decline from shorter to longer wavelengths.  We have adjusted a synthetic SED composed by linear combination of composite stellar populations (CSP) and a two parameters extinction law from Cardelli (1989). 
The base is composed by 6 CSP from Charlot \& Bruzual models (Charlot \& Bruzual 2007) with a Salpeter IMF (Salpeter 1955) and  a $\tau$-exponentially declining star formation history with $\tau=$100 Myr. According to the low metallicity detected in the gaseous phase we have selected CSP with sub-solar metallicity Z=0.5 Z$_\odot$. 
The ages of the 6 stellar population are 13 Myr, 200 Myr, 500 Myr, 1 Gyr, 4 Gyr and 7 Gyr, respectively. 
The results of the bestfit ( $\chi^2$=2.09) are shown in Table \ref{sed_fit}. We also present the median stellar population mass fraction for solution in the 68\% confidence interval. 
From the best fit, the stellar mass of the system is log M=10.00 $\pm$ 0.35 dex.

\begin{table*}
\caption[]{Fraction of different stellar populations for the best fit of the SED of J033227.07-274404.7 First two rows indicate the fraction of each composite stellar population (CSP) used for the fit. Last two rows show the fraction of stellar populations divided into young, intermediate and old.}
\centering
\begin{tabular}{l  c   c c    c c    r r   c c   c c c }
\hline
 & &\multicolumn{2}{c}{Young (Myr)} & & &\multicolumn{2}{c}{Intermediate (Gyr)} & & &\multicolumn{3}{c}{Old (Gyr)} \\
            & & 15       & 200  & & & 0.5  & 1    & & & 4    & & 7 \\
\hline
best fit    & & $<$ 0.01 & 0.16 & & & 0.08 & 0.05 & & & 0.05 & & 0.65 \\
median 68\% & & $<$ 0.01 & 0.15 & & & 0.22 & 0.07 & & & -- & 0.55 & -- \\
\hline
best fit    & &\multicolumn{2}{c}{0.17} & & &\multicolumn{2}{c}{0.13} & & &\multicolumn{3}{c}{0.70} \\ 
median 68\% & &\multicolumn{2}{c}{0.13 $\pm$ 0.07} & & &\multicolumn{2}{c}{0.32 $\pm$ 0.18} & & & \multicolumn{3}{c}{0.55 $\pm$ 0.21}  \\
\hline 
\end{tabular}
\label{sed_fit}
\end{table*}

\section{System configuration}
\label{conf}

\subsection{Velocity field and sigma map vs. observed morphology}
\label{int}

The composite image in the top left panel of Figure \ref{objeto_sims} shows an elongated structure about 14 kpc long. Given the arc-like nature of the feature, the possibility that we are observing a gravitational arc should be explored. Gravitational arcs were discovered more than twenty years ago (Soucail et al. 1987a). 
The origin of these structures is the gravitational lensing of light coming from a distant object by a dense mass distribution such as the core of galaxy clusters (Hammer \& Rigaut 1989; Estrada et al. 2007).
In order to determine if the structure seen is a gravitational arc, we analyzed the field of view (FoV) of  J033227.07-274404.7 in search of any mass distribution at a given redshift that might be deflecting the light from a distant object  (such as a cluster of galaxies).
We used the GOODS catalog to search around a radius of 180 arcsec\footnote{At the distance of the object, this equals to 1.3 Mpc} with a depth of 25 mag/arcsec$^2$ and found at most two galaxies sharing similar redshift ($z \pm 0.3 z$). This lack of galaxies sharing a given redshift can be interpreted as the absence of a particular mass distribution around the object (Soucail et al. 1987b), which rules out the possibility of a gravitational arc.

Once we have ruled out this possibility, we might suppose that the main, elongated structure is an inclined, distorted disk. If this were the case, the ellipticity of the main component of the system would equal $\sim$ 0.76, giving an inclination angle of 76 $\deg$, and implying the disk is close to edge-on.
Considering the extension of this disk ($\sim$14 kpc in diameter) and the mass derived from the SED fit, the disk would be rotating with a velocity, $V_{rot} \sim$ 80 km/s at 7 kpc from the center. Assuming the disk has an inclination of 76 $\deg$ and considering that the line-of-sight velocity, $V_{los}$, is given by  $V_{los} \ = \ V_{rot} \ \times \ \sin i$, where $i$ is the inclination of the system with respect to the plane of the sky, the gradient in the velocity field associated with the main structure of the system would be of approximately 80 km/s. 
\footnote{ In this calculation, stellar mass of the whole system is being used, resulting in a likely lower limit for $ V_{rot}$ and thus for  $V_{los}$.}

However, the velocity field derived from GIRAFFE observations (top left panel of Figure  \ref{objeto_sims}) shows a gradient smaller than 15 km/s within all the pixels associated with the main component of the system. A small gradient in the line-of-sight velocities in a disk galaxy is usually associated with an almost face-on system, where the contribution of the rotation velocity to the line-of-sight is small. In this case however, the morphology would seem to indicate that the system is almost edge-on in which case the line-of-sight velocity is similar to the rotation velocity of the system. So, what type of system are we actually observing?

The main clue is given by the VF. 
The fact that there is almost no velocity gradient for the main component of the system can only be explained if the galaxy is seen face-on. In which case, the bright, elongated structure seen in the direct images must be a bar lying on the disk of the main component. 
In this scenario, the disk would not be seen except for the faint structure seen to the southwest of the main galaxy. 
For bright barred galaxies in the Local Universe, the difference between the central magnitudes (both in the B and U bands) and the magnitudes of the outer parts of the disk can amount to 5 mag/arcsec$^2$ (Kuchinski et al. 2000).
The HST is only a 2.5 meters telescope and depth is an important issue for distant galaxies whose emissions are severely affected by cosmological dimming. For example, the optical radius (3.2 times the disk scale length) of a Milky Way like galaxy at z=0.5 requires 3 hours of HST-ACS observations to be properly recovered, and this is without accounting for extinction effects. 
Considering that the observed magnitude of the central parts of the system is 22.86 mag/arcsec$^2$ and that  the GOODS detection limit is about 25 mag/arcsec$^2$, this diluted disk would not be detected in the direct HST-ACS images.

The formation of this bright bar could have been triggered by the interaction with a companion.
This companion galaxy would be lying within the northern-most pixel of the IFU displaying emission. It would be orbiting around the main galaxy, moving with a velocity of at least 130 km/s. This velocity is estimated from the difference beween the average velocity of the main galaxy (-22 km/s) an the velocity given by the northern-most pixel (equal to 114  km/s). This velocity difference would imply that the companion galaxy orbit lies outside the plane of the main galaxy.
The $\sigma$ map of the system (top right panel of Figure \ref{objeto_sims}) shows a maximum of 48 km/s at the position between the companion and the northern tip of the main component, while the southern pixels associated with the main component show smaller values of $\sigma$ going from 31.4 km/s for the central parts of the galaxy to 19.50 km/s for the southern tip. The location of the maximum value of $\sigma$ could be pin-pointing the region in the system where the gas is most perturbed, indicating the interacting region between the main galaxy and the companion. 
If the small velocity gradient seen in the VF were the result of the superposition of several emission lines with different velocities for each IFU pixel (which would be the case if the galaxy had a large inclination value), then the  $\sigma$ map would display much larger values. See Figure 2 of Yang et al. (2008) for several examples regarding this situation.
Also in favour of this interacting scenario is the blue feature in the color image of the system (Figure \ref{color_objeto}) that could be associated with enhanced star formation associated with a tidal perturbation and thus be tracing the passage of the northern companion. 
The strong [NeIII] lines seen in the FORS2 spectrum could imply the presence of shocks produced by the interaction.  

Assuming there are two different interacting components, we estimated the mass ratio between them from their z band magnitudes. In order to do so,  we extracted the photometry in the z band for each object. This was done using the POLYPHOT task in IRAF\footnote{IRAF is distributed by the National Optical Astronomy Observatories, operated by the Association of Universities for Research in Astronomy, Inc., under cooperative agreement with the National Science Foundation.}. Sky was computed as the average of the sky in four regions around the object. Considering the same mass-to-light ratio for both components, we find a mass ratio of 5 between the main component and the companion \footnote{This value is actually a lower limit, since the companion is bluer than the main galaxy, as is shown in the color map (Fig. \ref{color_objeto}).}
.

\subsection{Numerical simulations of the encounter}
\label{num_sims}

In order to test the hypothesis of a face-on galaxy suffering an encounter with a less massive galaxy we ran several numerical simulations of plausible encounters using two different codes:
ZENO developed by Barnes\footnote{http://www.ifa.hawaii.edu/$\sim$barnes/software.html} (2002) and GADGET2 developed by Springel (2005). Both codes include standard N-body techniques for the collisionless components and Smooth Particle Hydrodynamics (SPH) for the gaseous component.

In the ZENO simulations, the galaxy models are similar to those of Barnes (1998, 2002): a bulge with a shallow cusp (Hernquist 1990), an exponential disk with constant scale height (Freeman 1970; Spitzer 1942) and a dark matter halo with a constant density core (Dehnen 1993; Tremaine et al. 1994).
The gas has the same properties as in Barnes (2002) and the SPH computation follows the isothermal equation of state.
For the GADGET2 simulations, the galaxies are constructed using a spherical dark matter halo (with a Hernquist profile, Hernquist 1990) which contains a disk, composed of stars and gas, and a bulge.  Galaxies are created following Springel et al. (2005).
Further details on the simulations using GADGET2 are given in Peirani et al. (2009).
For the orbital parameters, several configurations were explored with both numerical codes:  coplanar+direct, coplanar+retrograde, direct+slightly inclined orbits, and retrograde+slightly inclined orbit. All configurations followed parabolic orbits. Mass ratios of 3:1, 4:1 and 5:1 were tested.

The inclined and direct 3:1 mass ratio encounter as presented by Barnes (2002) 
seemed to best reproduce both the observed morphology and kinematics. 
For this particular simulation, the baryonic matter equals 20\% of the dark matter.  The gas is initially distributed like the disk component.

The total mass of the system (main galaxy and companion) was set to 1.6666 mass units.
In the ''ZENO'' code, G=1. 
The length unit was set to 95.3 kpc. 
Considering that the total stellar mass of the system derived from the SED fit equals 10$^{10} \ M_\odot$, the other code units are: 1.41 $\times$ 10$^{11} \ M_\odot$ for the mass, 1.17 Gyrs for the time and 80 km/s for the velocity. The simulation considered $N_{halo} + N_{stars} + N_{gas} = 17856 + 21744 + 23994 = 63594 $ particles.
The progenitor disk in the model deviates from the Tully-Fisher (TF) relation. When considering a disk that follows TF, the simulated bar is smaller than the observed one. Though the global morphology of the feature, as well as the velocities derived, follow the observations.

In order to properly compare the kinematics of the simulated encounter with the observed one, a grid ressembling the GIRAFFE IFU was superposed on the resulting snapshots. Velocities within each ''pixel'' of the grid were averaged in order to derive a velocity field similar to the one shown in the top middle panel in Figure \ref{objeto_sims}. 
Middle panel on the bottom row of Figure \ref{objeto_sims} shows a snapshot of the encounter at t=1.65 (1.93 Gyr) from the first peripassage and at t=0.01 (12.0 Myr) from the second one. 
Middle panel on the bottom row of the same Figure shows the simulated velocity field of the 
gas, right panel of the same row shows the velocity dispersion map as they would be observed with the IFU of GIRAFFE.
For both an inclined and retrograde 4:1 encounter and a 5:1 encounter, a thicker and rounder bar than the one seen in the observations is formed. In both these cases, the velocity field trend is reproduced but the $\sigma$ peak is misplaced with respect to the observations. For this reason, the 3:1 encounter was chosen.

\section{Discussion and Conclusions}
\label{disc}

The simulated velocity field (bottom middle panel of Figure \ref{objeto_sims}) follows the same trend as the observed one (top middle panel of the same Figure) though the difference between the northern-most pixel and the average value of what seems to be the main galaxy, i.e. the other five pixels, is higher. For the observation the difference is of 130 km/s while for the simulation this difference equals 169 km/s. However for the lower five pixels of the IFU, the velocity gradient of the simulation is also small. For the observations, the velocity gradient for these pixels equals 14 km/s, while for the simulations this gradient equals 23 km/s. 
For the sigma map, the distribution of the velocity dispersion in the simulated map (right bottom panel of Figure \ref{objeto_sims}) is similar to the observed one (top right panel of the same Figure). 
The largest sigma value for the simulation equals  96 km/s, while for the observations, this sigma peak has a value of 48 km/s. The average sigma value of the simulation (39 km/s) is close to that of the observations (31 km/s).
Globally the trends in both radial velocity and dispersion are followed. 
The observed morphology is also reproduced if cosmological dimming is taken into account.
Scaling the particle density of the simulation to the observed magnitude of the central parts of the system (22.86 mag/arcsec$^2$), a difference of 4 mag/arcsec$^2$ is seen between the core of the system and the fainter structures of the underlying disk. Considering that the GOODS detection limit is about 25 mag/arcsec$^2$, the outer parts of the disk would not be detected in the direct images.

If we now follow the evolution of the simulation, a bar begins to form about 200 Myr after the first peripassage. This bar starts to fade out before the second peripassage, but is enhanced again once this second peripassage occurs. During the encounter the disk is disrupted though not destroyed.
In order to look for any signs of a disrupted disk in our observations, we analyzed the raw FORS2 spectrum associated with the system  searching for any particular features that could be related to the small feature seen southwest from the main galaxy. This small ``blob'' is barely seen in the HST image on the top panel of Figure \ref{objeto_sims}, but can be seen in the color image in the bottom panel of the same Figure.
In the FORS2 spectra at the position of the main galaxy, a very slight asymmetry in the detected flux is seen. More flux seems to be detected on the side of the galaxy closer to the small ``blob''. This could be indicating that the ``blob'' has the same velocity as the main galaxy and thus belongs to its disk. Unfortunately the resolution and S/N of the raw spectra do not allow to draw a conclusive argument on this issue.

The fact that the mass ratio from the measured light is lower than 3:1 can be explained by tidal stripping of the smaller galaxy during the direct encounter. 
The slight discontinuites seen in the brightness of the HST composite image could be due to the presence of dust lanes. Notice how these features seem to be transverse to the elongated structure. If this galaxy were edge-on, one would expect the dust lanes to appear along the main axis of the galaxy, that is along one side and the other of the bar. Instead these features seem to lie perpendicular to this axis and may be realted to the presence of arms or rings on the plane of the galaxy.

\begin{figure*}
\centering      
\includegraphics[scale=0.4, angle=0]{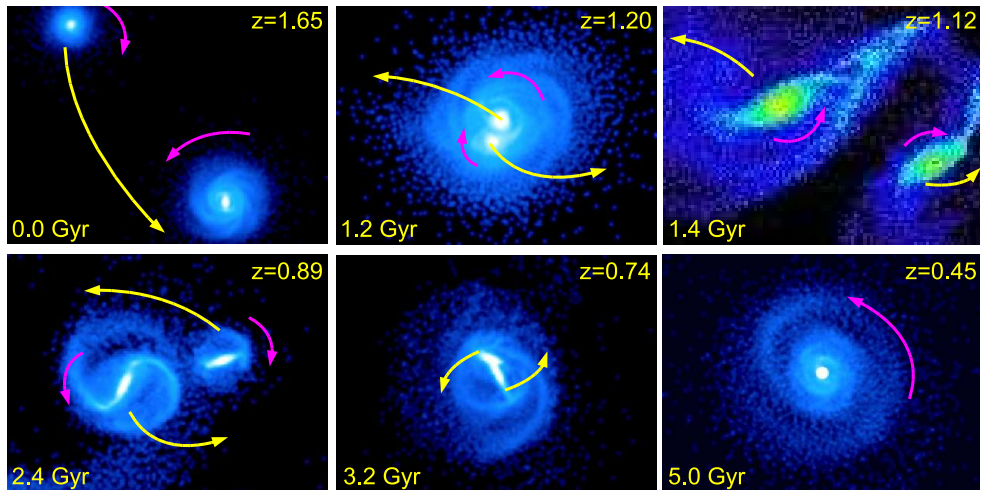}
\caption{{\it From left to right, top to bottom:} Time evolution of the projected gas density of the simulated encounter from the beginning of the simulation until 5 Gyr after. The first peripassage of the galaxies, a zoom of the encounter at the moment when the bar begins to form, the time when galaxies are most distant after the first peripassage and the second peripassage are shown.  Size of frame differs for each snapshot in order to show different stages of the encounter. The fifth snapshot in the sequence shows the frame that is currently being ''observed''.}
\label{evo_sim} 
\end{figure*}

\begin{figure}
\centering      
\includegraphics[scale=0.275, angle=0]{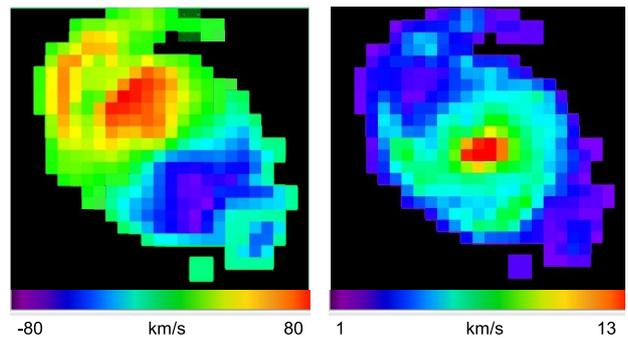}
\caption{ {\it Left:} Velocity field of the merger remnant 5 Gyrs after the beginning of the simulation -last snapshot shown in Figure \ref{evo_sim}. The velocity gradient associated with a slightly inclined rotating disk is seen. {\it Right:} $\sigma$ map of the same snap. Maximum value matches the center of the merger remnant.}
\label{kin_sim} 
\end{figure}

The simulation was left to evolve all the way to the post-merger stage. Figure \ref{evo_sim} shows the evolution of the simulated encounter from the beginning of the encounter to the end of the merging process. The second snapshot on the top row shows the moment of the first peripassage. A zoom of the central parts of the system when the bar begins to form can be seen in the third snapshot of the top row of the Figure. The second snapshot in the bottom row of the Figure corresponds to the second peripassage of the galaxies, just 12 Myr before the observed morphology and kinematics of the system are reproduced. 
Right-most snapshot in the bottom row of the Figure shows the outcome of the merger. Figure \ref{kin_sim} shows the VF and the $\sigma$ map computed for this snap. Resolution has been enhanced with respect to that of the GIRAFFE observations in order to be able to distinguish any particular velocity patterns.
The VF derived from the merger remnant of the simulation shows a gradient similar to that of a slightly inclined rotating disk with an amplitude of $\sim$ 160 km/s from one end of the galaxy to the other. The $\sigma$ peak matches the center of this disk as would be expected for a rotating disk.

Thus the outcome of this 3:1 merger seems to be a disk galaxy.
Notice that for this simulation, the fraction of gas considered at the beginning of the simulation equals 12\% of the mass of the disk -which is the gas fraction also used by Barnes in his simulations of mergers of gas-rich disk galaxies (Barnes 2002).
For the observed system, the gas fraction of the progenitor is much larger. This fraction can be computed following the method by Hammer et al. (2009) that takes into account the SED fit, the SFR of the system and the merger dynamical time -see Hammer et al. (2009, 2009b) and Puech et al. (2009) for further details.
Following this computation, the initial fraction of gas in the progenitor amounts to 60\% of the disk mass. 
In a previous work, Hopkins et al. (2009) have shown that if the gas fraction of the progenitor galaxy is 
lies between 60\%- 80\% of the disk mass, the disk will rebuild after an encounter, even if the mass ratio is 1:1.
So in the case of this system, for which the gas fraction is large and the mass ratio is 3:1, a disk would be the expected output after the merging process.
This can be seen from the simulations even if the gas fraction considered for the simulated progenitor is considerably lower.
In this case, the mass ratio and the configuration of the encounter are such that a disk is formed even if the gas fraction is relatively low as the simulations have shown.

We have presented results from multi-wavelength observations of  J033227.07-274404.7, a perturbed system at z=0.74  Using various techniques we have analysed the morphology, star-formation history and kinematics of this elongated, arc-like sytem. The analysis of the surroundings of this system have ruled out the possibility of a gravitational arc.
Though the morphology could indicate that this object is an almost edge-on galaxy, the kinematical information imposes very important constraints which have led us to explore other possible configurations. 
This system seems to be composed of an almost face-on disk galaxy and a second disk galaxy 3 times less massive following an inclined orbit. 
The presence of a rather young stellar population on both components suggests on-going enhanced star-formation due to the interacting process. 
Numerical simulations have been used to trace the history of the encounter. We seem to be witnessing the encounter at its second passage. 
The analysis of this system shows the necessity of combining both morphology, photometry, kinematics and numerical simulations in order to disentangle the complex systems seen at intermediate redshifts and to determine the outcome of their evolution.
This object is amongst the best example on how one may misinterpret morphology in absence of kinematical information.

\begin{acknowledgements}
IFC acknowledges the Sixth Program of the EU for a Marie Curie Fellowship, as well as the H.H. FC Foundation for their financial and moral support.      
\end{acknowledgements}

\end{document}